\documentclass[epsf]{aa}
\usepackage{natbib}
\usepackage{graphicx}

\begin{document}

\title{Two X-Ray Bright Cataclysmic Variables with Unusual Activities:
       GZ Cnc and NSV 10934}
\subtitle{}
\authorrunning{T. Kato et al.}
\titlerunning{Two Unusual X-Ray Bright Cataclysmic Variables}

\author{Taichi Kato\inst{1}
        \and Pavol A. Dubovsky\inst{2}
        \and Rod Stubbings\inst{3}
        \and Mike Simonsen\inst{4}
        \and Hitoshi Yamaoka\inst{5}
        \and Peter Nelson\inst{6}
        \and Berto Monard\inst{7}
        \and Andrew Peaece\inst{8}
        \and Gordon Garradd\inst{9}
}

\institute{
  Department of Astronomy, Kyoto University, Kyoto 606-8502, Japan
  \and MEDUZA group, Vedecko-kulturne centrum na Orave,
          027 42 Podbiel 194, Slovakia
  \and 19 Greenland Drive, Drouin 3818, Victoria, Australia
  \and 46394 Roanne Drive Macomb, MI USA 48044
  \and Faculty of Science, Kyushu University, Fukuoka 810-8560, Japan
  \and RMB 2493, Ellinbank 3820, Australia
  \and Bronberg Observatory, PO Box 11426, Tiegerpoort 0056, South Africa
  \and 32 Monash Ave, Nedlands, WA 6009, Australia
  \and PO Box 157, NSW 2340, Australia
}

\offprints{Taichi Kato, \\ e-mail: tkato@kusastro.kyoto-u.ac.jp}

\date{Received / accepted }

\abstract{
  We report on a discovery of unexpected activities in two X-ray bright
dwarf novae.  GZ Cnc showed an anomalous clustering of outbursts in
2002, in contrast to a low outburst frequency in the past record.
The activity resembles an increased activity seen in some intermediate
polars or candidates.
We identified NSV 10934, X-ray selected high-amplitude variable star,
as a dwarf nova with unusually rapid decline.  The outburst characteristics
make NSV 10934 a twin of recently discovered intermediate polar (HT Cam)
with dwarf nova-type outbursts.  We propose that these activities
in X-ray strong dwarf novae may be a previously overlooked manifestation
of outburst activities in magnetic cataclysmic variables.
\keywords{
Accretion, accretion disks --- novae, cataclysmic variables
           --- Stars: dwarf novae
           --- Stars: individual (GZ Cnc, NSV 10934)}
}

\maketitle

\section{Introduction}

   Cataclysmic variables (CVs) are close binary systems consisting of a
white dwarf and a red dwarf secondary transferring matter via the Roche-lobe
overflow.  Some of CVs show dwarf nova (DN) outbursts, which are believed
to be a consequence of the instabilities in accretion disks
\citep{osa96review}.  Outbursts of DNe usually occur semi-regularly.
The presence of unusual CVs with DN-like outbursts, which significantly
deviate from the canonical picture of DNe, has recently been receiving
special attention \citep{ish02htcam,szk02dodra}.  Some of these objects
have been proven to intermediate polars (IPs) [for recent reviews of
IPs, see \citet{pat94ipreview,hel96IPreview}], which contain weakly
magnetized white dwarfs.  Theoretical attempts have been also made to
explain these unusual outburst characters in the presence of a global
magnetic field (e.g. \cite{ang89DNoutburstmagnetic}).
We hereby report on the discovery
of two X-ray bright DNe, which may be further candidates for these
IPs with unusual DN-like outbursts.

\section{GZ Cnc}

   GZ Cnc was discovered as a variable star by Takamizawa (Tmz~V34).
Subsequent observations revealed that this object is a dwarf nova
which is identified with a ROSAT source \citep{kat01gzcnc}.
The object was independently confirmed to be a cataclysmic variable
in the course of optical identifications of ROSAT bright sources
\citep{bad98RASSID}.  \citet{jia00RASSCV} reported an optical spectrum
which showed strong Balmer and He\textsc{I} emission lines.  Although
\citet{jia00RASSCV} did not explicitly mentioned, He\textsc{II} emission
lines were detected stronger than in typical dwarf novae
\citep{wil83CVspec1}.

   \citet{kat01gzcnc} obtained time-resolved CCD photometry of the 2000
February long outburst.  The long duration of the outburst and the slow
rising rate suggested that GZ Cnc is a good candidate for a long-period
dwarf nova.  As reported in \citet{kat01gzcnc}, the recorded outbursts
up to 2001 were relatively rare.  Although there were unavoidable
seasonal observational gaps, only three outbursts were recorded between
1999 and 2001.

\begin{figure}
  \begin{center}
  \includegraphics[angle=0,height=5.5cm]{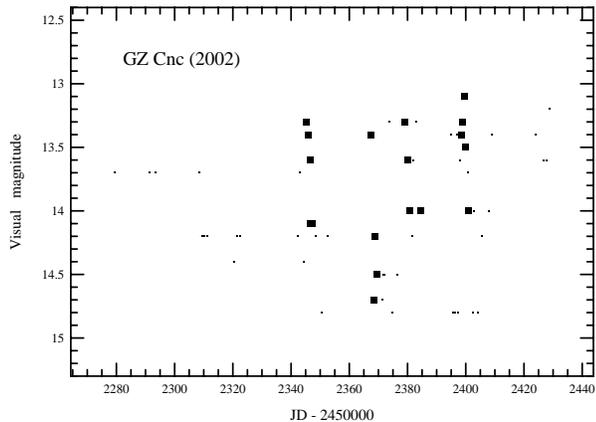}
  \end{center}
  \caption{Light curve of GZ Cnc in the 2002 season.  Large and small dots
  represent positive and negative (upper limit) observations, respectively.
  Note the high frequency of outbursts (large dots).}
  \label{fig:gzcnc1}
\end{figure}

   In 2002 March -- May, we noticed a dramatic increase of the outburst
frequency.  Fig. \ref{fig:gzcnc1} shows the light curve of the 2002 season,
mainly drawn from visual observations by the authors.  Some additional
observations (visual and CCD) reported to the VSNET Collaboration\footnote{
  http://www.kusastro.kyoto-u.ac.jp/vsnet/} have been incorporated.
All observers used $V$-band calibrated comparison stars.  The uncertainties
of the observations are 0.2--0.3 mag, which will not affect the following
discussion.

   Table \ref{tab:gzcncburst} lists the known outbursts of GZ Cnc since
the discovery by Takamizawa.
The shortest interval outbursts in the 2002 unusually active
season was only 11 d, and the other two intervals were 21--22 d.
As shown in Fig. \ref{fig:gzcnc1}, the durations of the outbursts in 2002
were very short in contrast to the long outburst in 2000 February
\citep{kat01gzcnc}.  Although such bimodal activity may suggest
an outburst activity seen in SU UMa-type dwarf novae
\citep{war85suuma}, the apparent lack of superhumps
during the long outburst seems to exclude the possibility of an SU UMa-type
dwarf nova \citep{kat01gzcnc}.

   Alternately, the present behavior in some aspects resembles a
``clustering" of outbursts observed in some IPs
[EX Hya: \citet{hel89exhya}; TV Col: Uemura et al. in preparation],
whose interpretation is still in debate \citep{hel00exhyaoutburst}.
Although no clear coherent pulses were detected during the 2000 February
outburst \citep{kat01gzcnc}, the presence of He\textsc{II} emission lines
and the relatively strong, hard X-ray spectrum \citep{bad98RASSID}
makes some resemblance to IPs.

\begin{table}
\caption{Outbursts of GZ Cnc.}\label{tab:gzcncburst}
\begin{center}
\begin{tabular}{ccccc}
\hline\hline
\multicolumn{3}{c}{Date} & JD-2400000 & Max \\
\hline
1994 & November & 30 & 49687 & 13.1$^a$ \\
2000 & February & 3  & 51578 & 13.7$^b$ \\
2000 & December & 29 & 51908 & 13.1 \\
2002 & March    & 11 & 52345 & 13.3 \\
2002 & April    & 2  & 52367 & 13.4 \\
2002 & April    & 14 & 52378 & 13.3 \\
2002 & May      & 4  & 52399 & 13.1 \\
\hline
 \multicolumn{5}{l}{$^a$ Discovery observation by Takamizawa.} \\
 \multicolumn{5}{l}{$^b$ Long outburst reported in \citet{kat01gzcnc}.} \\
\end{tabular}
\end{center}
\end{table}

\begin{figure}
  \begin{center}
  \includegraphics[angle=0,height=5.5cm]{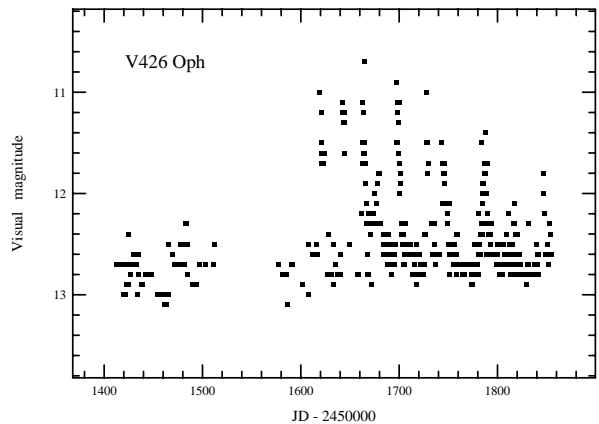}
  \end{center}
  \caption{An state of increased outburst activity in V426 Oph (1999--2000).
  The data are from reports to VSNET.}
  \label{fig:gzcnc2}
\end{figure}

   The present activity can also be comparable to V426 Oph, another
dwarf nova which is known to show occasionally increased activities
(Fig. \ref{fig:gzcnc2}; see also \cite{wen90v426oph}).
V426 Oph has been also suggested to be an IP
\citep{szk86v426ophEXOSAT}, although this possibility is recently questioned
\citep{hel90v426oph}.  The relatively hard X-ray spectrum of V426 Oph
\citep{ver97ROSAT} is also suggestive of an analogy between GZ Cnc and
V426 Oph, which may comprise a new class of cataclysmic variables with
prominent occasional increases of outburst activities.  These activities
may be a result of the weak presence magnetic fields, although the
evidence of the magnetic nature (at least in V426 Oph) is still
tantalizing.

\section{NSV 10934}

\begin{figure*}
  \begin{center}
  \includegraphics[angle=0,height=5.0cm]{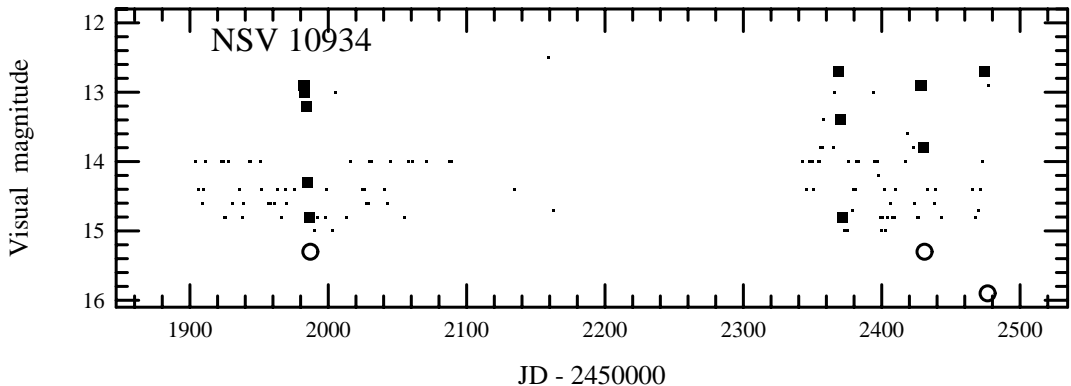}
  \end{center}
  \caption{Long-term light curve of NSV 10934.  Large and small dots
  represent positive and negative (upper limit) observations, respectively.
  Open circles are unfiltered CCD measurements, which have a sensitivity
  close to $R_{\rm c}$.  The most recent two CCD observations (NP and NP)
  have been calibrated by using GSC 9523.1025 (Tycho-2 magnitude: $V$ = 11.73,
  $B-V$ = +1.07).  The overall uncertainty of the CCD photometry is 0.2 mag.}
  \label{fig:nsv10934-1}
\end{figure*}

   NSV 10934 was discovered as a large-amplitude suspected variable star
of unknown classification.  The cataloged range of variability was
11.2 to [15.0 p.  We noticed that the object can be identified with
a bright ROSAT X-ray source (1RXS J184050.3$-$834305).  Since a combination
of a large-amplitude variation and the strong X-ray emission suggests
a cataclysmic variable, we started systematic monitoring of this
NSV object through the VSNET Collaboration (vsnet-chat 3340).\footnote{
  http://www.kusastro.kyoto-u.ac.jp/vsnet/Mail/chat3000/\\msg00340.html
}  The first outburst was detected on 2001 March 13 by RS (vsnet-alert
5778).\footnote{
  http://www.kusastro.kyoto-u.ac.jp/vsnet/Mail/alert5000/\\msg00778.html
}  Three additional outbursts have been recorded since 2002 July.
The well-observed most recent two outbursts have been characterized
by a sudden rise (more than 1.3 mag within 1 d), which established the
dwarf nova-type variability.  Table \ref{tab:nsv10934burst}
lists the observed outbursts.
Fig. \ref{fig:nsv10934-1} depicts the long-term light curve based
on visual observations by the authors (RS, NP, PA) and snapshot CCD
observations.  The accuracy of the visual observations is 0.2--0.3 mag,
which will not affect the following discussion.
The shortest interval of the observed outbursts is 46 d.

\begin{figure}
  \begin{center}
  \includegraphics[angle=0,width=4.3cm]{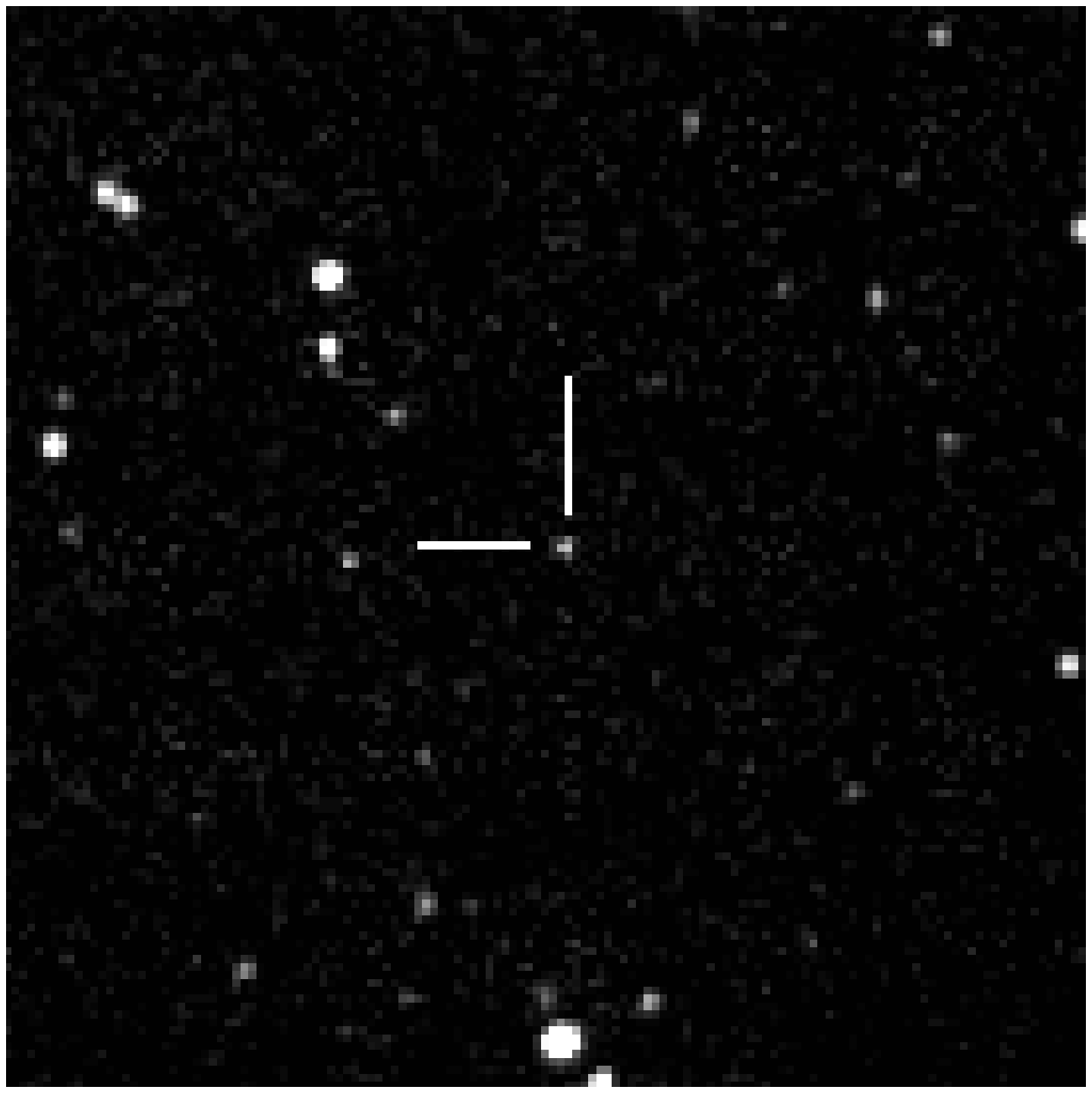}
  \includegraphics[angle=0,width=4.3cm]{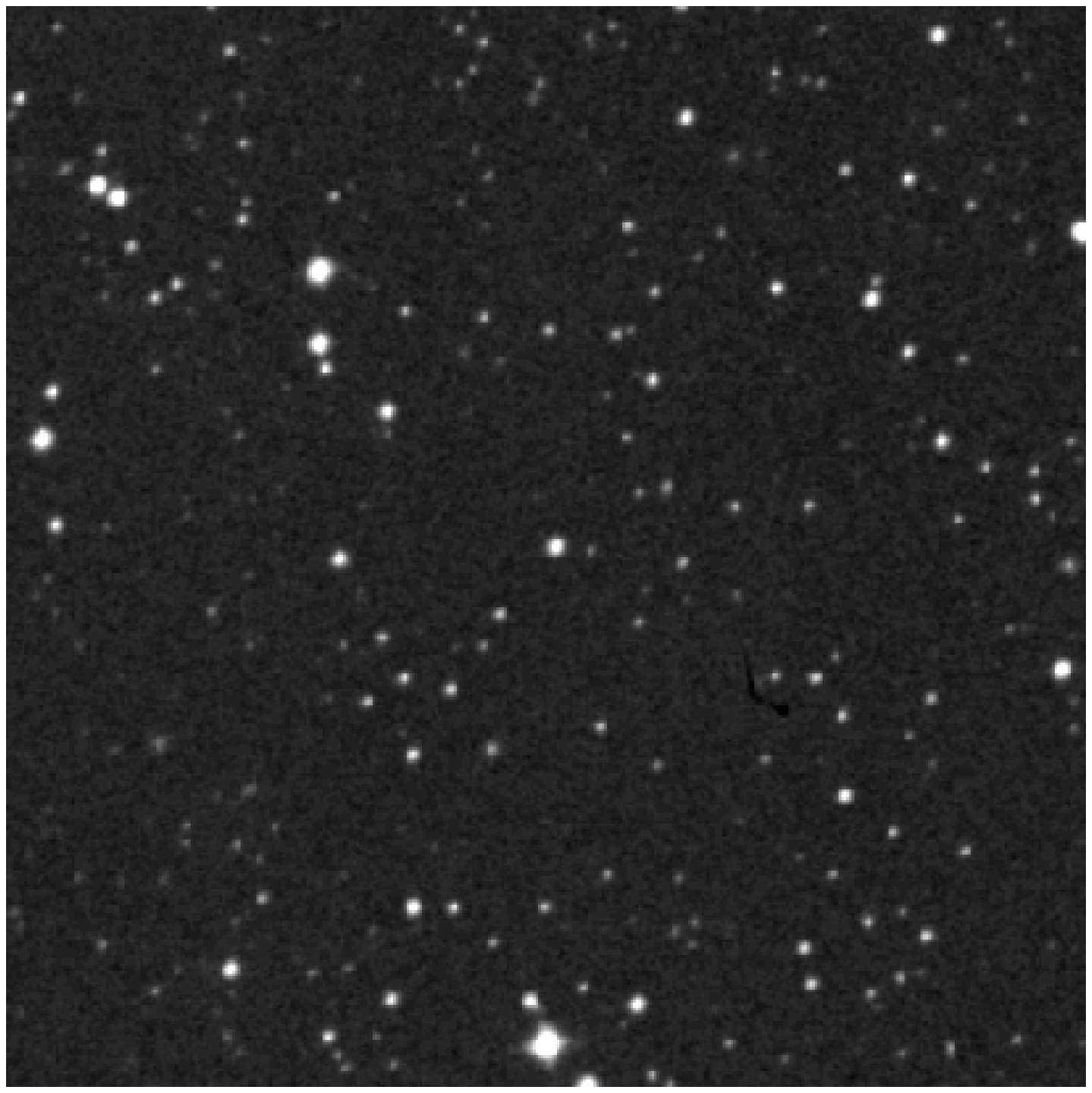}
  \end{center}
  \caption{Identification of NSV 10934.  (Left) PN's image on 2002
  June 5.423 UT (8 arcminutes square, north is up, and east is left;
  magnitude 15.3, slightly above quiescence).  (Right) DSS red image showing
  NSV 10934 in quiescence.}
  \label{fig:nsv10934-2}
\end{figure}

   Astrometry of NSV 10934 was performed on CCD images taken by PN
(2002 June 5.423 UT) and BM (2002 July 21.024 UT),
both of which were taken during the rapid decline stage from outbursts.
The variability of the object has been confirmed by a comparison between
the two images.  An average of measurements of two images
(UCAC1 system, 182 and 71 reference stars respectingly; internal
dispersion of the measurements was $\sim$0$''$.1) has yielded
a position of 18$^h$ 40$^m$ 52$^s$.52, $-$83$^{\circ}$ 43$'$ 09$''$.84
(J2000.0).  The position agrees with the USNO$-$A2.0 star at
18$^h$ 40$^m$ 52$^s$.28, $-$83$^{\circ}$ 43$'$ 09$''$.2
(epoch 1983.040 and magnitudes $r$ = 15.6, $b$ = 16.5), or the
GSC-2.2.1 star with position end figures of 52$^s$.420 and 09$''$.74 
(epoch 1993.767 and magnitudes $r$ = 15.11, $b$ = 17.06),
which is most likely the quiescent counterpart of NSV 10934
(Fig. \ref{fig:nsv10934-2}).

\begin{figure}
  \begin{center}
  \includegraphics[angle=0,height=6cm]{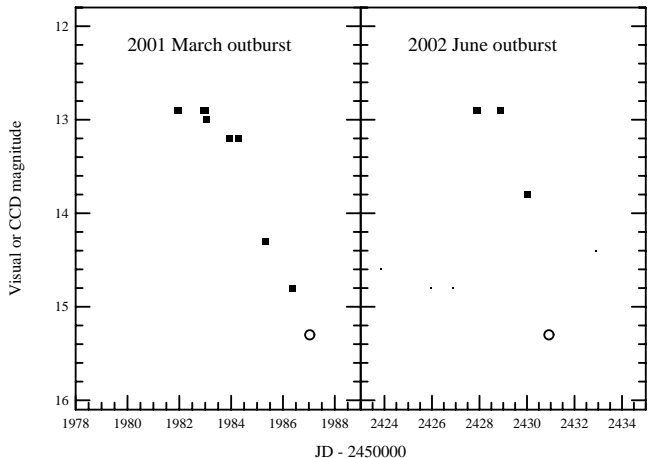}
  \end{center}
  \caption{Enlarged light curve of the best observed outbursts of NSV 10934.
  The symbols are the same as in Fig. \ref{fig:nsv10934-1}.}
  \label{fig:nsv10934-3}
\end{figure}

   Fig. \ref{fig:nsv10934-3} shows the enlarged light curve of the best
observed outbursts.
All the recorded outbursts rather quickly faded.  The most recent two
outbursts faded more than 1 mag within 3 d of the outburst maximum.
Linear fits to the best-observed decline stages of the first two
outbursts have yielded decline rates of 0.71$\pm$0.06 mag d$^{-1}$ and
0.71$\pm$0.03 mag d$^{-1}$, respectively.  Rather fragmentary data of
the recent two outbursts further suggest an even higher value close
to the termination of the outbursts: the object faded by 1.5 mag in 0.90 d
(JD 2452430.02--.92) and by 3.2 mag in 2.59 d (JD 2452473.93--76.52).

\begin{table}
\caption{Outbursts of NSV 10934.}\label{tab:nsv10934burst}
\begin{center}
\begin{tabular}{ccccc}
\hline\hline
\multicolumn{3}{c}{Date} & JD-2400000 & Max \\
\hline
2001 & March & 13 & 51982 & 12.9 \\
2002 & April & 4  & 52369 & 12.7 \\
2002 & June  & 2  & 52428 & 12.9 \\
2002 & July  & 18 & 52474 & 12.7 \\
\hline
\end{tabular}
\end{center}
\end{table}

   The combination of relatively strong and relatively hard (see Table
\ref{tab:rosat}) X-ray detection and short optical outbursts
either suggests the possibility of an IP with dwarf nova-like outbursts,
a non-magnetic HT Cas-like unusual dwarf nova with rather irregularly
spaced short outbursts \citep{kat02ircom,woo95htcasXray}, or a system
resembling an unusual dwarf nova BZ UMa with short outbursts and
quasi-periodic oscillations \citep{kat99bzuma,jur94bzuma}.

   The precipitous fading (1.5 mag in 0.90 d) recorded during the terminal
stage of the outbursts is unlike usual dwarf novae.  The sequence of
a more slowly fading plateau phase near outburst maxima and a subsequent
rapid fading more resembles the behavior of an outburst in the recently
discovered IP, HT Cam \citep{ish02htcam}.  The X-ray hardness ratios are
also similar (Table \ref{tab:rosat}).  HT Cam showed a gradual decline
for the first 0.5 d, followed by a dramatic decline by more than 4
mag d$^{-1}$ \citep{ish02htcam}.  Since the time-evolution of the light
curve is slightly slower in NSV 10934, the orbital period of NSV 10934
is expected to be slightly longer than that of HT Cam (86 min), if
NSV 10934 indeed turns out to be an HT Cam-like object.

\section{Conclusion}

   We report on a discovery of unexpected activities in two X-ray bright
dwarf novae.  GZ Cnc showed an anomalous clustering of outbursts in
2002, in contrast to a low outburst frequency in the past record.
The activity resembles an increased activity seen in some intermediate
polars or candidates.
We have shown that the outburst characteristics of NSV 10934
closely resembles those of recently discovered intermediate polar (HT Cam)
with dwarf nova-type outbursts.
The X-ray properties of these objects are summarized in
Table \ref{tab:rosat}.  We propose that these activities
in X-ray strong dwarf novae may be a previously overlooked manifestation
of outburst activities in magnetic cataclysmic variables.
Further research to elucidate the relation between
these unusual cataclysmic variables and IPs is encouraged.
Although direct detection of IP pulses is known to be sometimes difficult
or tantalizing (see \citet{ros94v426ophswumav348pup} for a recent example)
we encourage more extensive search for the IP-type coherent signal in
quiescence and outburst (cd. HT Cam: \cite{kem02htcam,ish02htcam}),
and in X-rays.

\begin{table}
\caption{Comparison of X-ray Properties of GZ Cnc, NSV 10934 and HT Cam$^a$.}
         \label{tab:rosat}
\begin{center}
\begin{tabular}{ccccc}
\hline\hline
Object    & Count rate & HR1 & HR2 & $V$ \\
\hline
GZ Cnc    & 0.181 & 0.53 & 0.15 & 15.4 \\
NSV 10934 & 0.239 & 1.00 & 0.50 & 15.9 \\
HT Cam    & 0.152 & 0.79 & 0.43 & 16.2 \\
\hline
 \multicolumn{5}{l}{$^a$ The X-ray data are taken from \citet{ROSATRXP}.} \\
\end{tabular}
\end{center}
\end{table}

\vskip 1.5mm

This work is partly supported by a grant-in aid [13640239 (TK),
14740131 (HY)] from the Japanese Ministry of Education, Culture, Sports,
Science and Technology.
The CCD operation of the Bronberg Observatory is partly sponsored by
the Center for Backyard Astrophysics.
The CCD operation by PN is on loan from the AAVSO,
funded by the Curry Foundation.
This research has made use of the Digitized Sky Survey producted by STScI, 
the ESO Skycat tool, the VizieR catalogue access tool.

\end{document}